\documentstyle[12pt]{article}
\def\baselinestretch{1.5}
\begin{document}
\thispagestyle{empty}
\null\vskip -1cm
\centerline{
\vbox{
\hbox{April 2, 1997}\vskip -9pt
\hbox{hep-ph/9704219}\vskip -9pt
     }
\hfill 
\vbox{
\hbox{UICHEP-TH/97-2}
     }     } \vskip 1cm

\begin{center}\large \bf 
CP Violation and Scalar Lepton \\
Flavor Oscillation  \\
\end{center}
\vspace{1cm}
\centerline{David Bowser-Chao and Wai-Yee Keung}
\begin{center}
\it 
Physics Department, 
University of Illinois at Chicago, 
IL 60607-7059, USA\\
\end{center}
\vspace{.5cm}
\begin{abstract}
Lepton flavor violation can be induced in supersymmetry by
the mixing of two or more of the leptonic scalar partners, 
$\tilde l \rightarrow \tilde l'  ,$ 
where $\tilde l, \tilde l'= \tilde e, \tilde\mu, \tilde\tau.$ 
Krashnikov and Arkani-Hamed {\it et al.} pointed out that 
this effect may be observable
at the Next Linear Collider, through slepton pair
production with subsequent lepton flavor violating decays.
If the slepton mixing involves all three generations, 
CP violation from the Cabibbo-Kobayashi-Maskawa phase
could lead to asymmetries between the observed LFV decays
$\tilde l \rightarrow \tilde l'$ and
$\tilde{\bar l}\rightarrow \tilde{\bar l'}$.
We lay down the formalism and give simple expressions
for the CP violating asymmetry in the transition probabilities,
and consider possible signals at future colliders.
\end{abstract}

\vspace{1in}
\centerline{PACS numbers:  
11.30.Er, 11.30.Hv,  14.80.Ly, 14.60.Gh
\hfill}
\newpage

\section*{Introduction}

\quad\vskip -1cm

Quark flavor is not conserved in nature, as manifestly
demonstrated by
oscillation in the $K$--$\bar K$ system. There is no
evidence to date, however, for the analogous lepton flavor
violation. The branching ratio for $\mu \to e\gamma$, for example,
is less than $4.9\times 10^{-11}$ at the $95\%$ confidence level\cite{pdata}.
This is well in accord with the Standard Model,
where  lepton flavor violation (LFV) is expected to be
highly suppressed, due to  the smallness of $m_{\nu}/v$, where $m_{\nu}$ is 
the largest neutrino mass and $v$ the electroweak breaking scale.

Lepton flavor violation is also suppressed in exactly supersymmetric extensions
of the Standard Model; supersymmetry, however, is necessarily broken, and soft
symmetry breaking terms can in  general generate significant  LFV\cite{susylfv}.
Recently, Krashnikov\cite{krash} and Arkani-Hamed\cite{Hall} {\it et al.} showed
that the resultant mixing between at least two of the charged slepton generations
could produce dramatic lepton flavor violating signals at LEPII or the Next
Linear Collider (NLC). While detailing the particular case of two generation
mixing, these authors also noted the obvious extension to mixing amongst all
three generations.

Beyond simply complicating the analysis of lepton flavor violation, moving to
three generation mixing also gives rise to a new source of CP violation (CPV);
namely, the single complex phase in the Cabibbo-Kobayashi-Maskawa (CKM) matrix
that relates the electroweak flavor basis to the three mass eigenstates. In this
paper we examine the consequences of this CP violating phase\cite{pil}.

The analysis is similar in spirit to CPV from three family neutrino
oscillation\cite{Barger}. We shall provide 
compact formulas for the CPV transition probability asymmetry
$P(\tilde{\bar l} \rightarrow \bar l') -P(\tilde l \rightarrow
l')$,
where $\tilde l,\tilde l'$
denote the sleptons $\tilde e,\tilde\mu,\tilde\tau$ 
and 
$\tilde{\bar l},\tilde{\bar l'}$ are the respective anti-particles.
Following Ref.\cite{Barger}, we show that the CPV result is identical
for the available choices
$(l,l')=(e,\mu),(\mu,\tau),$ and $(\tau,e)$\ ,
then consider observation of the asymmetry at future colliders. 

\newpage
\section*{Formalism}

\quad\vskip -1cm

First, we shall assume  (as did Refs. \cite{krash,Hall}) (1) the presence of 
LFV soft symmetry breaking terms in the electroweak scale
lagrangian, without regard for the nature of their origin, and
(2) that left-right slepton mixing is small. The second assumption
simplifies analysis of CPV and LPV effects, which then have separate
contributions from the left and right-handed sleptons. This assumption
is not strongly model-dependent, since left-right slepton mixing
is generally suppressed by the ratio of the lepton mass to the left-right slepton
mass splitting, with the latter set by the $D$-term to be of order the
electroweak scale\cite{gab}. Furthermore, we shall see below that CPV effects 
involving left-right mixing are further suppressed by the the ratio of the
slepton width to the left-right mass splitting. We shall thus follow Refs. \cite{krash,Hall} 
and focus on the right-handed sleptons, while noting that generalization of 
our analysis to the separate left-handed slepton contribution to CPV is also possible.

Due to mixing, the flavor states ($\tilde e$, $\tilde \mu$, $\tilde \tau$) do not
evolve in time with a trivial phase. We define the mixing
amplitudes $U_{\alpha i}$,  where the flavor state is expressed by
superposition of mass eigenstates $|S_i\rangle$ of masses $m_i$,
\begin{equation}
| \tilde l_\alpha\rangle  = \sum_i U_{\alpha i}|S_i\rangle 
\ ,\quad
| S_i\rangle  = \sum_i U_{\alpha i}^*|\tilde l_\alpha\rangle 
\ .
\end{equation}
The matrix $U$ can be explicitly written using the 
standard KM parameterization with three mixing angles $\theta_i$ 
and one phase $\delta$,
\begin{equation}
\left(
      \begin{array}{c} 
        \tilde{e} \\ \tilde{\mu} \\ \tilde{\tau} \\
      \end{array}
\right)=
\left(
    \begin{array}{ccc}
     c_1       &  s_1 c_3                    & s_1 s_3 \\
    -s_1c_2    &  c_1c_2c_3+s_2s_3e^{i\delta} & c_1c_2s_3-s_2c_3e^{i\delta}\\
    -s_1s_2    &  c_1s_2c_3-c_2s_3e^{i\delta} & c_1s_2s_3+c_2c_3e^{i\delta}\\
    \end{array}
\right)
\left(
      \begin{array}{c} 
        S_1 \\ S_2 \\ S_3 \\
      \end{array}
\right) \ ,
\end{equation}
with $s_i=\sin\theta_i$ and
$c_i=\cos\theta_i$.

If the LSP is a neutralino $\chi^0$ (which we assume to be primarily Bino), 
the flavor eigenstate 
$l$ can be tagged by the decay $\tilde l \to l \chi^0$.
The flavor state $|\tilde\alpha\rangle$ 
$(\alpha=e,\mu,$ or $\tau)$ of a slepton
produced at $t=0$ in the rest frame,
could be resolved as $|\tilde\beta\rangle$ at a later time $t$ 
with the transition amplitude,
\begin{equation}
A(\tilde{\alpha}\rightarrow\tilde{\beta};t)=\sum_i U_{\alpha i}U^*_{\beta i}
\exp(-i m_i t-\hbox{$1\over2$} \Gamma t)
\ .
\end{equation}
For simplicity we assume this to be the only open decay channel; if not, 
results below should be multiplied by the decay branching ratios\cite{Hall}.
We also take all states to have a common decay width $\Gamma$ and 
\begin{equation}
\Delta m_{ij} \equiv m_i - m_j\ ;\  
\Delta m_{ij},\Gamma \ll \bar{m}
\equiv \hbox{$1\over3$}(m_1+m_2+m_3)
\ .
\label{eq:condition}
\end{equation}
Further restrictions on $\Delta m_{ij}$ are discussed below.

The time-dependent oscillation probability is
$p(\tilde \alpha\rightarrow \beta;t) 
= |A(\tilde \alpha\rightarrow \tilde{\beta} ;t)|^2 $; note that we omit 
the decay product neutralino in our notation.
The time--averaged probability that is actually measured is given by
\begin{equation}
P(\tilde \alpha\rightarrow\beta)=\int_0^\infty 
p(\tilde \alpha\rightarrow\tilde{\beta};t)dt{\Large /}\int_0^\infty
\sum_\beta p(\tilde \alpha\rightarrow\tilde{\beta};t)dt
\label{eq:transition}
\end{equation}
$$
=U_{\alpha i}U_{\beta i}^* U^*_{\alpha j} U_{\beta j} 
 \left( {\Gamma\over\Gamma+(m_i-m_j)i} \right) 
=P(\tilde{\bar\beta}\rightarrow\bar{\alpha}) \ .
$$
From the unitarity of $U_{\alpha i}$ and the second line of the equation
above, it can be seen that lepton flavor violation vanishes\cite{Hall} as
$\Delta m_{ij}\ll \Gamma$; in the limit of exactly degenerate
sleptons, the mass basis can be arbitrarily rotated to bring $U_{\alpha i}$
into diagonal form.

CP violation is reflected in asymmetry between the transition
probabilities of CP--conju-gate channels, 
$P(\alpha\rightarrow\beta)-P(\tilde{\bar\alpha}\rightarrow\bar\beta)$.
Since
$P(\tilde{\bar\alpha}\rightarrow\bar\beta)
=P(\tilde \beta\rightarrow\alpha)$ 
from CPT invariance, the difference flips
sign under interchange of  $\alpha$ and $\beta$ and vanishes for the
diagonal case $\alpha=\beta$. 
It is convenient to introduce the antisymmetric symbol 
$\varepsilon_{\alpha\beta}$ such that 
$\varepsilon_{e\mu}=\varepsilon_{\mu\tau}=\varepsilon_{\tau e}=1$.
We summarize all properties of asymmetry,
\begin{equation}
   P(\tilde      \alpha     \rightarrow      \beta) 
  -P(\tilde{\bar \alpha   } \rightarrow \bar  \beta)    
=\varepsilon_{\alpha\beta} {\cal A}
\ ,
\quad 
{\cal A}
 = - 4\ \hbox{Im}(U_{e1}U^*_{\mu 1}U^*_{e2}U_{\mu2})
     \ \hbox{Im}(\phi_{12}+\phi_{23}+\phi_{31})
\ ,
\label{eq:asym}
\end{equation}
where
\begin{equation}
\phi_{ij}={\Gamma\over \Gamma+(m_i-m_j)i}\ ,\quad
\hbox{Im} \phi_{ij}={(m_j-m_i)\Gamma \over \Gamma^2+(m_i-m_j)^2} \ .
\end{equation}
From Eq.(\ref{eq:condition}), this asymmetry
result could also  have been obtained by using the narrow width approximation.
Note that $\cal A$ vanishes not just  for $\Delta m_{ij} \ll \Gamma$ as discussed
above, but also if $\Delta m_{ij}\gg \Gamma$,  since in this limit
$\hbox{Im} \phi_{ij}$ vanishes as well. This is simply due to the fact that
CP violation requires interference between $U_{\alpha i}$ and the final state
decay phases in in Eq.(\ref{eq:transition}); in the limit of very large 
slepton mass splitting, Eq.(\ref{eq:transition}) becomes an incoherent sum over 
the mass eigenstates.
For CP violation from slepton
mixing, we thus require $\Delta m_{ij} \sim \Gamma$. We can also see that
CPV effects  involving left-right mixing are indeed suppressed by the the ratio of the
slepton width to the left-right mass splitting.

The expression for $\cal A$ involves the single universal CPV
combination\cite{Chau} of mixing amplitudes in the case of three families,
\begin{equation}
X_{\rm CP} \equiv
\hbox{Im}(U_{e1}U^*_{\mu 1}U^*_{e2}U_{\mu2})=-s_1^2c_1s_2c_2s_3c_3\sin\delta
\ .
\end{equation}
The magnitude of this CPV parameter $X_{\rm CP}$ ranges from zero to
the maximum value $\sqrt{3}/18=0.096$. Ref.\cite{Hall} shows 
that the present experimental bound on
$\hbox{Br}(\mu\rightarrow e\gamma)$ puts no significant constraint on
the mixing angles, with $(\Delta m/\bar m) \sin2\theta
\stackrel{<}{\sim} 0.01$.

One may wonder how the experimental limit\cite{EDM} on the electric
dipole moment (EDM) of the electron constrains the phase in $U$. It
turns out that at the one loop level, the EDM requires both the left 
and right handed sleptons to participate in the loop.  Each
handedness contributes a GIM-like suppression factor, $\Delta
m_{L,R}/\bar m_{L,R} (\approx 0.01$ for our case). Together with a
factor $m_l/{\bar m}$ from the soft SUSY-breaking $A_l$ term or the supersymmetric
$\mu$ term, which mixes the $L$ and $R$ components, there is sufficient
suppression to expect no significant constraint on the
phase (note that the electron EDM {\em does} constrain the phase in the
left-right slepton mixing\cite{susylfv}, as well as putting tight limits
on a relative phase between the complex 
mass of the gauge fermion and the $A_l$ or $\mu$ term\cite{Nilles}).

\section*{Experiments}

\quad\vskip -1cm

It is believed that the cleanest environment to search for 
sleptons is at the proposed Next Linear $e^+e^-$ collider (NLC) at an
energy of $\sqrt{s} \simeq $ 500 GeV to 1.5 TeV. The signature for
pair production of sleptons is, assuming the neutralino decay discussed above,
a pair of charged leptons plus missing energy.  In
Fig.~1, we illustrate the cross section  for right handed
slepton production. At $\sqrt{s}=500$ GeV,
$\sigma(\tilde{e}\tilde{\bar e})$ is almost as large as 1 pb under the
assumption that the neutralino exchanged in the $t$-channel is
purely Bino with mass $M_\chi=50$ GeV (with an obvious degradation for an
admixture of Bino). Ref. \cite{Hall} calculated the 
backgrounds from $WW$, $e^\pm\nu W^\mp$, and $(e^+e^-)W^+W^-$ to total
about 12 pb. Through efficient cuts\cite{Becker}, Ref. \cite{Hall} estimates
a reduction of the background to about 5 fb for unpolarized beams, with
30\% signal acceptance, with further improvement possible with right-handed
polarized beams. Because the CPV signals considered here are simply asymmetries
in the LFV signals of the sort considered in Ref. \cite{Hall}, we shall simply
rely on these figures to provide an estimate of the sensitivity to 
CP violation from slepton mixing.

We start with a simple CP-odd observable which ignores the full
correlation between the lepton $l$ and the anti-lepton
$\bar{l'}$. From the sample of dilepton events that pass the cuts, we consider
the CPV asymmetry between muon and anti-muon events,
\begin{equation}
S_\mu ={N(\mu^-)-N(\mu^+)\over N(\hbox{all  signal}) }
\ .
\label{eq:S}
\end{equation}
Clearly only unequal flavor dilepton ($\mu^\pm l^\mp, l\ne\mu$)
events contribute to this asymmetry.

Since $S_\mu$ only involves a single particle count, it can be expressed 
directly in
terms of the single particle transition probability (Eq.(\ref{eq:transition})), 
\begin{equation}
S_\mu={\sigma(\tilde{e}\tilde{\bar e})
    [P(\tilde{e}\rightarrow \mu)   -P(\tilde{\bar e}\rightarrow \bar\mu)]
   +\sigma(\tilde{\tau}\tilde{\bar \tau})
    [P(\tilde{\tau}\rightarrow \mu)-P(\tilde{\bar \tau}\rightarrow \bar\mu)]
 \over
   \sigma(\tilde{e}   \tilde{\bar  e  })
  +\sigma(\tilde{\mu }\tilde{\bar \mu })
  +\sigma(\tilde{\tau}\tilde{\bar \tau})
  } \ .
\end{equation}
Using Eq. (\ref{eq:asym}), we obtain
\begin{equation}
S_\mu
={\cal A}f\ , \quad f\equiv
  {\sigma(\tilde{e}\tilde{\bar e})
  -\sigma(\tilde{\tau}\tilde{\bar \tau})
  \over
   \sigma(\tilde{e}   \tilde{\bar  e  })
  +\sigma(\tilde{\mu }\tilde{\bar \mu })
  +\sigma(\tilde{\tau}\tilde{\bar \tau})
  }  \ .
\label{eq:asym_nlc}
\end{equation}
At the tree level, $\sigma(\tilde{\tau}\tilde{\bar \tau})$ and
$\sigma(\tilde{\mu}\tilde{\bar \mu})$ come only from the $\gamma$-$Z$
exchange amplitude in the $s$ channel, but $ \sigma(\tilde{e}
\tilde{\bar e })$ involves additional neutralino exchange amplitude in
the $t$-channel. Thus $f$ in Eq. (\ref{eq:asym_nlc}) is generally non-zero.
The fraction $f$  can be close to one in the
limit that the $t$-channel diagram dominates, which will of course be true
given sufficiently high collider energy.
Fig.~1 shows $f$ to be 0.75 at $\sqrt{s}=500$ GeV and $\bar m$=150 GeV,
under the assumption that the exchange gaugino is purely Bino with mass
$M_\chi=50$ GeV.  
With an integrated luminosity of 100 fb$^{-1}$, around at
least $10^4$ events from the slepton pair production will
survive the dedicated cuts\cite{Hall,Becker}  which severely reduce the predictable
backgrounds to only about 500 $\mu^\mp e^\pm$ events
and about the same number of $\mu^\mp \tau^\pm$ events. 
Thus, it is possible to measure
the CP asymmetry $S_\mu$ at the level $0.02$ with 5$\sigma$.

The other flavor asymmetries are $S_e=0$ (which vanishes because
$\sigma(\tilde{\tau}\tilde{\bar \tau})=\sigma(\tilde{\mu}\tilde{\bar\mu})$)
and
$S_\mu=-S_\tau$. Statistics will be doubled if we take the combination
$S_\mu-S_\tau$.

In Fig.~2, we show  the size of $S_\mu-S_\tau$ 
versus $\Delta m/\Gamma$ for $f=0.75$, in the scenario that the mass differences
are equal in magnitude to the width, namely,
$m_1-m_2=m_2-m_3=\Delta m \sim \Gamma$.

In the  scenario where $ \sigma(\tilde{e} \tilde{\bar e }) 
\approx      \sigma(\tilde{\mu}\tilde{\bar \mu })  
\approx      \sigma(\tilde{\tau}\tilde{\bar \tau }) $ (e.g., if the
$t$-channel gaugino mass is relatively high), 
we can no longer use the uncorrelated asymmetry
$S$ to probe CPV.
We can, however, turn to other correlated observables such as 
$S_{\mu\bar e}$, which is defined by replacing the numerator in Eq.(\ref{eq:S}) 
by $N(\mu^- e^+)-N(\mu^+ e^-)$.

The overall amplitude for tagging $\mu \bar e $ at time $t$ and $\bar t$
after the production comprises two contributions. %
One is the $t$-channel gaugino exchange amplitude (for the {\it
selectron} production only) which has the time evolving factor
\begin{equation}
A({\tilde e}\rightarrow \mu) A(\bar {\tilde e}\rightarrow\bar e)
=\sum_i U_{e i}U^*_{\mu i}
\exp(-i m_i t-\hbox{$1\over2$} \Gamma t)
\sum_j U^*_{e j}U_{e j}
\exp(-i m_j \bar{t}-\hbox{$1\over2$} \Gamma \bar t) \ .
\label{eq:t}
\end{equation}
The other is the common $s$-channel $\gamma$--$Z$ exchange amplitude 
which has a simpler factor due
to unitarity of $U$ matrix,
\begin{equation}
\sum_\alpha 
A({\tilde \alpha}\rightarrow \mu) A(\bar {\tilde \alpha}\rightarrow\bar e)
=
\sum_i U_{\mu i}^*U_{e i}\exp[(-im_i-\hbox{$1\over2$}\Gamma)(t+\bar t)]
\ .
\label{eq:s}
\end{equation}
Denoting the asymmetry $S^0$ in the special case that the $s$ channel dominates, 
we have  the simple expression,
\begin{equation}
S^0_{\mu\bar e}
=-\hbox{$1\over3$}\cdot
4 X_{\rm CP}\ \hbox{Im} (\phi^2_{12}+\phi^2_{23}+\phi^2_{32})
=-S^0_{\mu\bar\tau}
=S^0_{e\bar\tau}
\ .
\label{eq:blind}
\end{equation}
The statistics factor $1\over3$ can be offset by summing all three
flavor asymmetries, as in 
$S^0_{\hbox{tot}} \equiv S^0_{\mu\bar e}-S^0_{\mu\bar\tau}+S^0_{e\bar\tau}$.

The asymmetry formula in the general case, where both ($s$ and $t$) channels  are 
important, is lengthy but quite straightforward, and which we present next.

The amplitudes for $e^-e^+ \rightarrow \tilde l
\ \tilde{\bar l}$ consist of a common element $M_0$ from the $s$-channel
$\gamma$--$Z$ exchange and an additional piece $M'$ for the case
$l=e$ from the $t$-channel gaugino exchange. Their explicit expressions
at the tree level are provided below.
From Eqs. (\ref{eq:t},\ref{eq:s}), we derive
\begin{equation}
{ d\sigma(e^-e^+ \rightarrow \tilde l\ \tilde{\bar l}\rightarrow
\alpha\bar{\beta} +\hbox{missing neutralinos})
\over 
d\cos\theta}=
 K \biggl( |M_0|^2\sum_{i,j} 
U_{\alpha i}^* U_{\beta i} U_{\alpha j} U_{\beta j}^* \phi_{ij}^2
\label{eq:dsig}
\end{equation}
$$
+|M'|^2 P(\tilde      e \rightarrow       \alpha) 
       P(\tilde{\bar e}\rightarrow {\bar\beta}) 
+2 \hbox{Re}(
M_0 M'^* \sum_{ijk} U_{\alpha i}^* U_{\beta i}
         U_{e k}^* U_{\alpha k} U_{e j} U_{\beta j}^*
\phi_{ik}\phi_{ij} ) \biggr)
\ .
$$
The flux and phase space factor $K$ is given below. 
The non-zero tree level  amplitudes $M_0$ and $M'$ for different
channels are\cite{Nojiri}
\begin{equation}
M_0(e^-_L e_R^+ \rightarrow \tilde l_R\ \tilde{\bar l}_R)
=e^2s \lambda^{1\over2}\sin\theta
\left( {1\over s} + {-{1\over2}+x_W \over (1-x_W)(s-M_Z)} \right) \ ,
\end{equation}
\begin{equation}
M_0(e^-_R e^+_L \rightarrow \tilde l_R\ \tilde{\bar l}_R)
=e^2s \lambda^{1\over2}\sin\theta
\left( {1\over s} + { x_W \over (1-x_W)(s-M_Z)} \right)  \ ,
\end{equation}
\begin{equation}
M'(e^-_R e^+_L \rightarrow \tilde e_R\ \tilde{\bar e}_R)
={e^2 s \lambda^{1\over2} \sin\theta \over (1-x_W)}
\sum_i {|V_{Bi}|^2\over t-M_{\chi_i}^2}   \ .
\end{equation}
Here, $\theta$ is the CM polar scattering angle, $M_{\chi_i}$ are the
neutralino masses ($i=1$ is the lightest), while $|V_{Bi}|^2$ is the
mixing probability of the Bino component in the $i$-th neutralino;
$\lambda^{1\over2} = \sqrt {1-4{\bar m}^2 /s}\;$ is the slepton
CM velocity, $s$ and $t$ are the usual invariant squares of energy and
momentum transfer, $x_W$ is the electroweak parameter,
$M_W^2/M_Z^2=1-x_W$. The flux and phase space factor  is 
$K=\lambda^{1\over2}/(128\pi s)$.

The  differential rate difference between   CP conjugate channels is
\begin{equation}
{d \Delta R_{\alpha\bar{\beta}} \over d\cos\theta} = 
 K \biggl(  |M_0|^2 \hat S^0_{\alpha\bar\beta}
+   |M'|^2\left(P(\tilde{e}\rightarrow \beta) \varepsilon_{e\alpha} 
               -P(\tilde{e}\rightarrow \alpha)\varepsilon_{e\beta}
          \right) {\cal A} 
\end{equation}
$$
-4
M_0 M' \sum_{ijk} \hbox{Im} 
        (U_{\alpha i}^* U_{\beta i}
         U_{e k}^*      U_{\alpha k} U_{e j} U_{\beta j}^*)
                    \hbox{Im}(\phi_{ik}\phi_{ij} ) \biggr)
\ .
$$
The asymmetry is
\begin{equation}
S_{\alpha{\bar\beta}}={
\Delta R_{\alpha\bar{\beta}}
\over
    \sigma(\tilde{e}   \tilde{\bar  e  })
  +\sigma(\tilde{\mu }\tilde{\bar \mu })
  +\sigma(\tilde{\tau}\tilde{\bar \tau})} \ ,
\label{eq:Semu}
\end{equation}
where the summed differential slepton cross-section follows trivially from
Eq. (\ref{eq:dsig}),
$$
d\sigma(\tilde{e}   \tilde{\bar  e  })
  +d\sigma(\tilde{\mu }\tilde{\bar \mu })
  +d\sigma(\tilde{\tau}\tilde{\bar \tau}) =     K\left( 3|M_0|^2+|M'|^2+2M_0M' \right) d\cos\theta \ .
$$
As expected, setting $\alpha=\mu$ and summing $\beta$  reproduces
the uncorrelated asymmetry in Eq.(\ref{eq:asym_nlc}).
Fig. 3 shows a comparison of CP conjugate events $\mu^-e^+$ (solid)
and $\mu^+ e^-$ (dashed) due to slepton pair production at
the NLC with $\sqrt{s}=500$ GeV.

\section*{Discussion}

\quad\vskip -1cm

Our study can be generalized to other processes. In the future Large
Hadron Collider (LHC), gluinos will be copiously produced, if the gluino
mass is about a few hundred GeV. Through the chain of cascade decays,
sleptons may occur in the intermediate state and give dilepton
events.  With enough CPV slepton flavor oscillation,
asymmetries of the type discussed here might be observable.
In the case of single gluino production, the single
lepton asymmetry in Eq. (\ref{eq:asym}) applies, where $\tilde\alpha$
denotes the first slepton produced in the decay chain, and $\tilde\beta$
is the second slepton. To avoid washing out of the asymmetry, some reconstruction
of the cascade decay would be necessary to at least statistically
identify the primary and secondary lepton decay products.

We also note that it is possible the NLC could be operated in a $\gamma\gamma$
mode\cite{gamgam}. In this case, all single lepton asymmetries $S_l$ vanish, but
$S_{\alpha\bar\beta}$ could be measured, and is given by Eq. (\ref{eq:blind})
or Eq. (\ref{eq:Semu})
with $M'$ set to zero. The signal and background rates for slepton pair
production are similar to those for intermediate mass charged Higgs production,
which are roughly equal after mild cuts\cite{bowser}. Both the
$t$-channel dominated limit of the $e^+e^-$ collider asymmetry $S_{\alpha}$ and the 
dilepton asymmetry measured by a $\gamma\gamma$ collider are proportional to $X_{CP}$,
the respective mass-splitting factors are  $\hbox{Im}(\phi_{12}+\phi_{23}+\phi_{31})$
and $\hbox{Im} (\phi^2_{12}+\phi^2_{23}+\phi^2_{32})$. Measurement
of both asymmetries could help indirectly measure the slepton mass-splitting.
Finally, we note that if  left-handed
polarization of the electron beam is possible, the $e^+e^-$ collider
measurement of $S_{\alpha\bar\beta}$ would yield the same information as that of the
$\gamma\gamma$ collider, since
$M'(e^-_L e_R^+ \rightarrow \tilde l_R\ \tilde{\bar l}_R)$ vanishes
at the tree-level.

\section*{Acknowledgment}

\quad\vskip -1cm

This work was supported in part by U.S.
Department of Energy under Grant number DE-FG02-84ER40173.

\thispagestyle{empty}
\null\vskip -1cm

\section*{Figure Captions}

\begin{itemize}

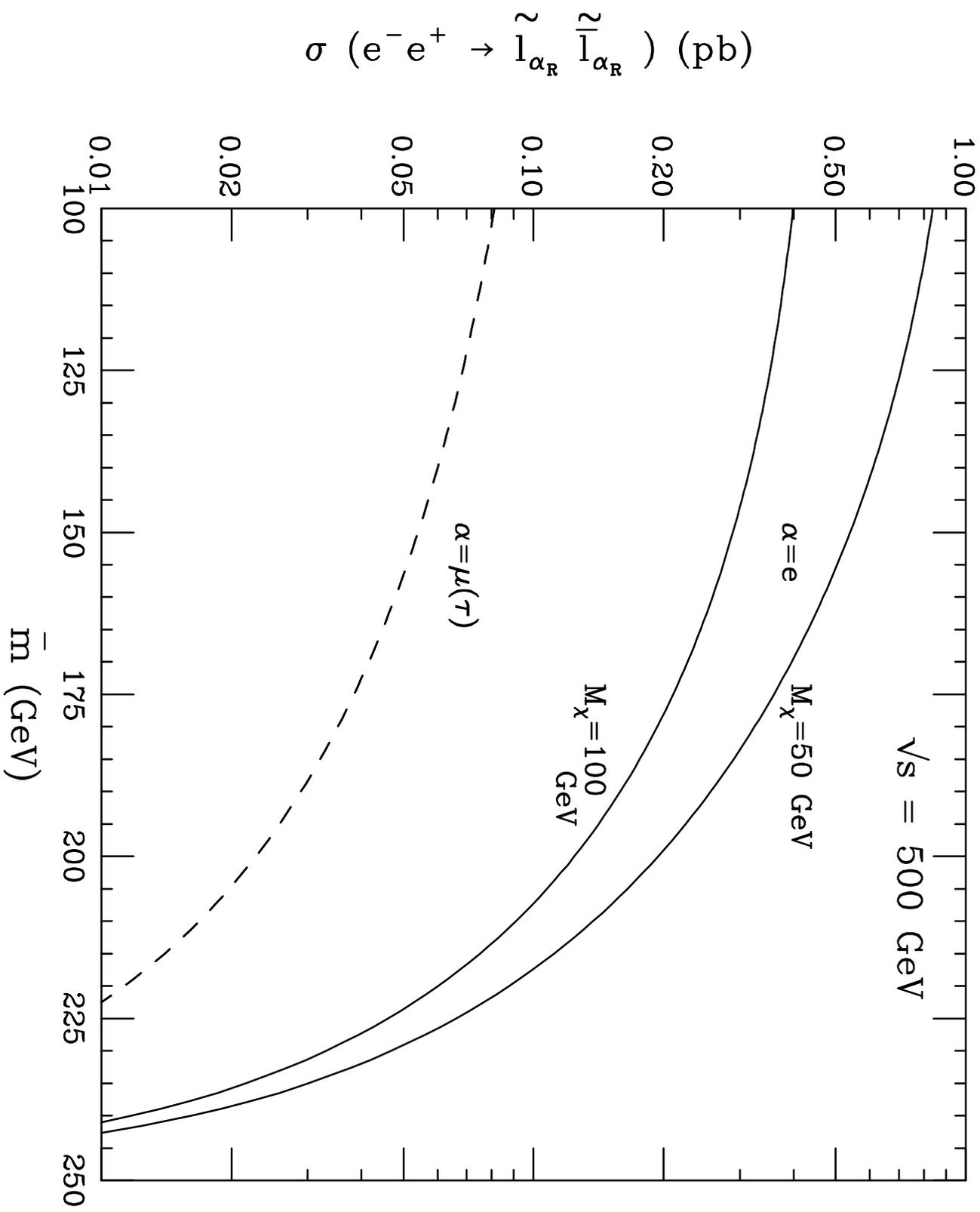
\item[Fig.~1] Dependence of the slepton pair production cross-section
on the slepton mass $\bar m$,
at the Next Linear $e^+e^-$ Collider, with $\sqrt{s}=500$ GeV.
The production of the selectron pair $\tilde{e}_R \tilde{\bar e}_R$
(solid curves) 
depends on the neutralino couplings and masses. Here we assume the
lightest neutralino of mass $M_\chi$ is purely Bino.  
We also show 
$\sigma(e^-e^+\rightarrow \tilde\mu_R \tilde{\bar\mu}_R)$ (dashed).
Note that $\sigma(e^-e^+\rightarrow \tilde\mu_R \tilde{\bar \mu}_R)=
           \sigma(e^-e^+\rightarrow \tilde\tau_R\tilde{\bar \tau}_R)$.

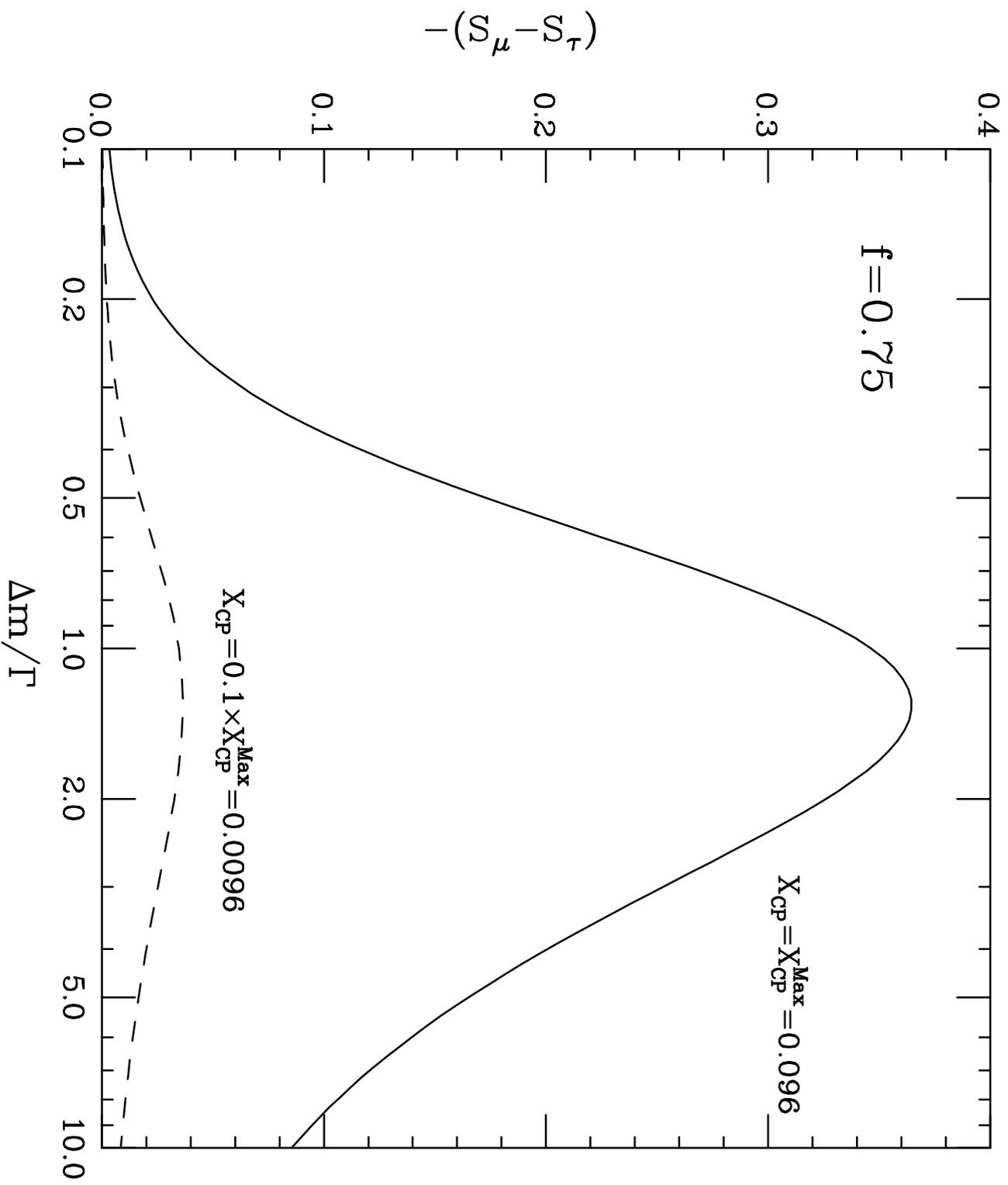
\item[Fig.~2] Uncorrelated asymmetry $-(S_\mu-S_\tau)$ versus 
$\Delta m/\Gamma$ for  $X_{\rm CP}=X_{\rm CP}^{\rm Max}$ (solid)  or\\
$X_{\rm CP}^{\rm Max}=0.1\times X_{\rm CP}^{\rm Max}$
(dashed), for $f=0.75$ and $\Delta m= \Delta m_{12} =\Delta m_{23}$.

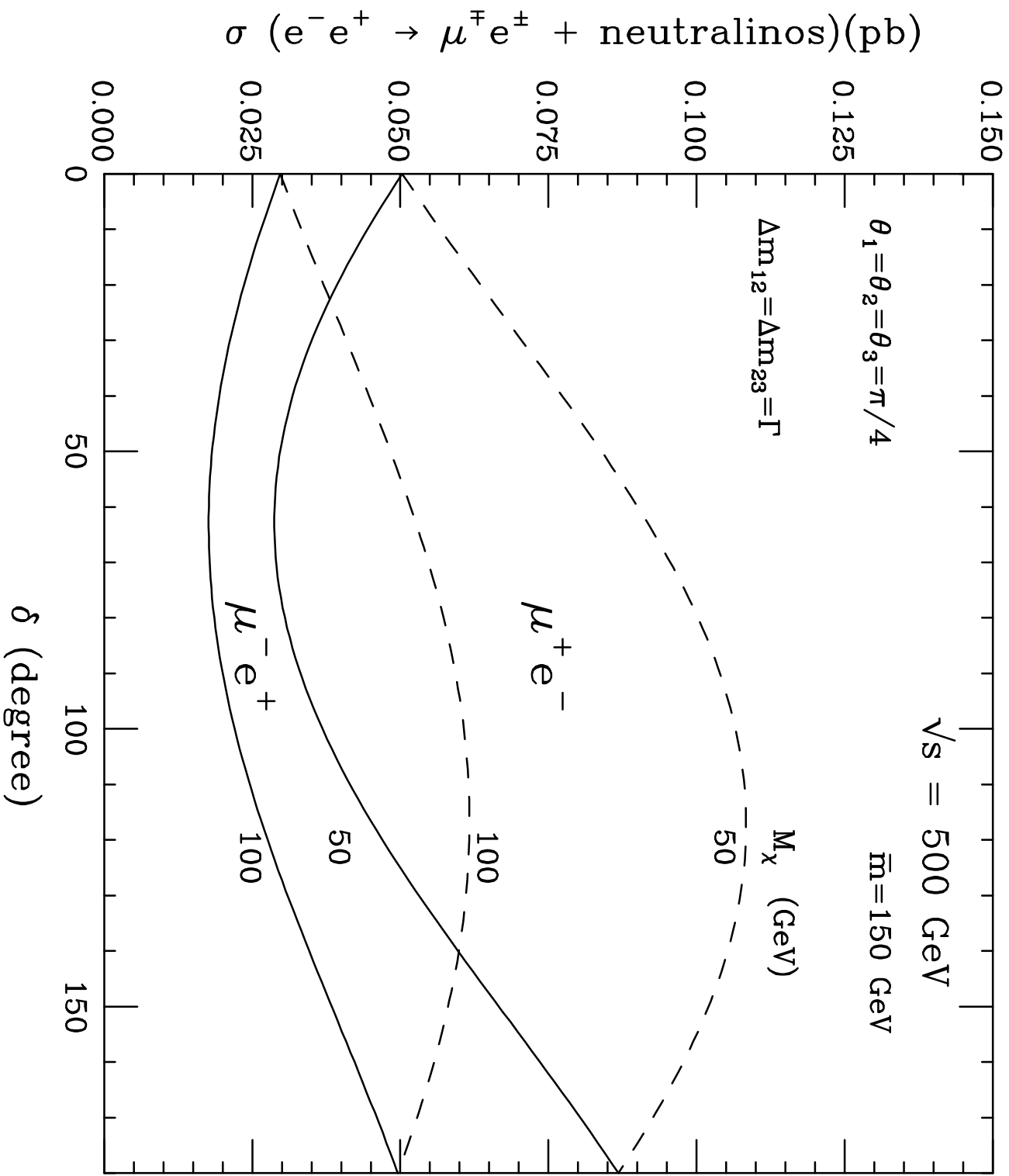
\item[Fig.~3] Comparison of CP conjugate events $\mu^-e^+$ (solid)
and $\mu^+ e^-$ (dashed) due to slepton pair production at
the NLC with $\sqrt{s}=500$ GeV. We choose  the case of maximal
mixings, $\theta_1=\theta_2=\theta_3={\pi\over4}$ and 
$\Delta m_{12}=\Delta m_{23}=\Gamma$. 
We illustrate the scenarios of $M_\chi=50$ or 100 GeV for ${\bar m}=150$
GeV. Event rates are shown versus the KM phase $\delta$. 
For $-\pi < \delta<0$, the event rates can be read by replacing $\delta$
by $|\delta|$ and reversing the labels between $\mu^-e^+$ and $\mu^+e^-$.

\end{itemize}
\newpage
\noindent 
{\it Note added:} While preparing this manuscript, we learned\cite{Feng}
that the authors of Ref. \cite{Hall} were extending their work to include
analysis of lepton flavor violation in the  three-generation case, 
as well as the CPV effects considered here\cite{followup}. 

\def\baselinestretch{1.2}

\end{document}